\documentclass[11pt, onecolumn]{article}

\usepackage[square,numbers,sort&compress,comma]{natbib}
\usepackage{url}
\usepackage{amsmath}
\usepackage{amssymb}
\usepackage{caption}
\usepackage{graphicx}
\usepackage{latexsym}
\usepackage{times}
\usepackage[pagewise]{lineno}
\usepackage{float}
\usepackage{bm}
\usepackage{booktabs}
\usepackage{mhchem}
\usepackage{lineno}
\usepackage{setspace}

\topmargin - 12pt 
\renewenvironment{abstract}%
              {
               \small
               {\bfseries \abstractname}
               \par
               \vspace{10pt}
              }

\renewcommand\abstractname{Abstract}

\newcommand{\nomenclature}
              [1]
              {
               \bgroup
               \flushleft
               \small\bf
               #1
               \par
               \egroup
              }

\renewcommand{\section}
              [1]
              {
               \bgroup
               \flushleft
               \small\bf
               \refstepcounter{section}
               \arabic{section}. #1
               \par
               \egroup
              }

\renewcommand{\subsection}
              [1]
              {
               \bgroup
               \flushleft
               \small\em
               \refstepcounter{subsection}
               \arabic{section}.
               \arabic{subsection}. #1
               \par
               \egroup
              }

\renewcommand{\subsubsection}
              [1]
              {
               \bgroup
               \flushleft
               \small\em
               \refstepcounter{subsubsection}
               \arabic{section}.
               \arabic{subsection}.
               \arabic{subsubsection}. #1
               \par
               \egroup
              }

  \newcommand{\acknowledgement}
              [1]
              {
               \bgroup
               \flushleft
               \small\bf
               #1
               \par
               \egroup
              }

  \newcommand{\sectionbib}
              [1]
              {
               \bgroup
               \flushleft
               \small\bf
               #1
               \par
               \egroup
              }

\setdisplayskipstretch{0.6} 

\begin{document}


\title{\LARGE Quantized Skeletal Learning (QSL): A Differentiable Programming Approach for Skeletal Reduction of Chemical Mechanisms}

\author{{\large Opeoluwa Owoyele}\\[10pt]
        {\footnotesize \em Department of Mechanical and Industrial Engineering, Louisiana State University, Baton Rouge, LA 70810, USA}\\[-5pt]}

\date{}


\small


\maketitle
\vspace{10pt}
\rule{\textwidth}{0.5pt}

\begin{abstract} 
\begin{spacing}{1.2}
  This paper presents a data-driven approach, referred to as Quantized Skeletal Learning (QSL), for generating skeletal mechanisms. The approach has two key components: (1) a weight vector that can be used to eliminate relatively unimportant species and reactions, and (2) an end-to-end differentiable program whose loss-function gradients, with respect to the weight vector, can be used to adjust those weights. To promote sparsity in the weight vector -- and to reduce the influence of certain reactions or species to zero -- an $l_1$-regularized objective is employed alongside the standard mean squared error loss, thus removing the least important components. The proposed QSL approach is validated by generating skeletal mechanisms for methane and ethylene based on the GRI 3.0 and USC II mechanisms, respectively, demonstrating effectiveness in deriving skeletal mechanisms with various levels of fidelity. Two variants of QSL, designated as QSL-R and QSL-S, are tested; these focus on eliminating reactions and species, respectively. Analysis of ignition delay times and species mass fractions demonstrate QSL's capabilities to reliably and efficiently extract data-driven skeletal mechanisms of varying fidelities from detailed mechanisms.  
\end{spacing}

\end{abstract}
\vspace{10pt}
\begin{spacing}{1.2}
\parbox{1.0\textwidth}{\footnotesize {\em Keywords:} skeletal mechanism generation; differentiable programming; automatic differentiation; data-driven reduction; chemical kinetics}
\end{spacing}
\rule{\textwidth}{0.5pt}
\vspace{10pt}

\clearpage

\vspace{20pt} 






\vspace{10pt}


\clearpage

\section{Introduction\label{sec:introduction}} \addvspace{10pt}
\begin{spacing}{1.2}
There is a continuing need to develop efficient ways to reduce chemical mechanisms, driven by the extreme computational costs of capturing chemical kinetics in numerical modeling of combustion \cite{lu2009toward}. Computational fluid dynamics (CFD) has played an increasingly significant role in recent decades in improving our understanding of combustion and aiding in the design of new combustors \cite{trendsturbulent}. However, the computational cost remains a significant issue \cite{pope2013small}. The sizes of chemical mechanisms vary widely, ranging from $\mathcal{O}$(10) reactions for hydrogen-air mechanisms to more complex hydrocarbon fuels that can contain as many as $\mathcal{O}$(1,000) species and $\mathcal{O}$(1,000--10,000) reactions \cite{lu2009toward, curran2019developing}. Even for the smaller hydrogen-air mechanisms, the majority of the computational time spent by a CFD solver is on solving the chemistry \cite{babkovskaia2011high}. Consequently, reducing the chemical complexity can result in a nearly proportional reduction in overall runtime. Over the past decades, several approaches for developing skeletal mechanisms have been devised, aimed at reducing the number of species or the number of reactions to make such models more feasible for CFD simulations.

The Directed Relation Graph (DRG) method \cite{lu2005directed, lu2006linear} uses a graph of interconnected nodes with connection weights between species pairs to identify and remove unimportant species from a chemical mechanism. This approach establishes a set of target species and specifies a predetermined allowable error threshold, $\epsilon$, then proceeds to remove species based on their impact on existing species within the graph. Since the introduction of the DRG method, several variants have been developed. For example, \cite{luo2010reduced} introduced a method for reducing mechanisms with several isomers by using the maximum norm instead of summation when computing connection weights. In DRG Restart \cite{lu2006systematic}, the DRG process is repeated a specified number of times, and changes to the graph across different trials can lead to potentially smaller skeletal mechanisms. Flux-based DRG \cite{tosatto2011transport} incorporates the effects of transport terms, such as those from neighboring grid cells. DRG with Error Propagation (DRGEP) \cite{pepiot2008efficient} allows errors to be damped as they propagate, enabling a finer selection of important chemical paths compared to the original DRG approach.

Other methods based on sensitivity analysis \cite{tomlin2013role, vom2019sensitivity, nouri2022skeletal} compute the changes in an output of interest (e.g., ignition delay, species concentrations) with respect to small changes in kinetic parameters. These methods often apply principal component analysis, either with respect to the final concentrations of important species \cite{vajda1985principal, turanyi1990reduction, xu1999simplification} or the net rates of species production \cite{turanyi1989reaction, borger1992extended, zsely2003influence, bahlouli2014reduced}. Sensitivities in these methods can be computed by solving a sensitivity equation, an adjoint equation, or using finite difference approximations.

In addition, optimization-based methods have been developed, where an objective function is minimized subject to certain constraints. Among these, methods using integer programming or genetic algorithms have been introduced \cite{petzold1999model, edwards2000reaction, androulakis2000kinetic, banerjee2003development, mitsos2008optimal, elliott2005reaction, elliott2006reaction}. For example, Baranwal et al. \cite{baranwal2024spin} proposed a Wiener filter-based system and mixed integer programming to minimize the number of retained reactions. Similarly, Fang et al. \cite{fang2025data} employed a sparse learning algorithm with an $\ell$-1 weight regularization to eliminate relatively unimportant reactions. In both cases, the reduction endeavor is treated as a \textit{static regression} problem. That is, the reduction is performed using species concentrations or reaction rates precomputed at certain time-stamps, without integrating the governing equations or differentiating through time.

In this work, we develop a new Quantized Skeletal Learning (QSL) algorithm based on the principles of differentiable programming. In this approach, a weight vector, $\bm{\alpha}$, is defined over either reactions or species concentrations and optimized through end-to-end differentiable integration of the governing equations. Regularization terms are incorporated into the objective function to promote sparsity, therby allowing for controlled trade-offs between mechanism compactness and predictive accuracy. The entire methodology is implemented in JAX \cite{jax2018github}, which allows efficient computation of exact gradients of the objective function via automatic differentiation.

Unlike recent data-driven reduction approaches (such as SPIN \cite{baranwal2024spin} and the sparse-learning framework of Fang et al. \cite{fang2025data}) that treat mechanism reduction as a static regression problem using precomputed time-series data, QSL explicitly integrates the chemical system and backpropagates through the ODE solver. This makes every component of the solver -- from chemical source-term evaluation to numerical time integration -- differentiable. Consequently, the optimization process captures the full temporal evolution of the system, rather than fitting isolated reaction-rate or limited concentration snapshots. This trajectory-level formulation allows QSL to account for the accumulation of integration errors and enforces dynamical consistency across the entire solution trajectory and under varying initial conditions. These capabilities are absent in static, non-differentiable approaches.

Also, While most existing methods focus on either eliminating species or reactions, QSL is flexible and can be applied to both. The current work involves the definition of a weight vector, $\bm \alpha$, applied to either the reactions or species concentrations and optimized via batch gradient descent. In the present study, QSL is first applied to reduce the number of species and reactions in GRI-Mech 3.0 \cite{gri_mech3}, targeting accurate prediction of ignition delays and transient or spatial species profiles. It is then extended to extract a skeletal mechanism for ethylene from the detailed USC Mech II mechanism \cite{usc_mechII}. In all cases tested, regularization terms are used to encourage the elimination of species or reactions. Mechanisms of varying levels of fidelity are produced by adjusting the amount of regularization applied to the primary objective function.


The remainder of this paper is structured as follows. In Section 2, an overview of the proposed approach is presented, detailing the structure of the QSL algorithm and the regularization terms used. In Section 3, results are shown for developing skeletal mechanisms using two different variants of the QSL approach, one targeting the elimination of reactions, and the second, the removal of species. The section demonstrates QSL's ability to develop mechanisms of varying fidelities, with extensions to generating mechanisms with NO chemistry. Finally, the paper concludes in Section 4 with some concluding remarks.

\section{Description of Quantized Learning Approach\label{sec:ql_approach}} \addvspace{10pt}
\subsection{Chemical reaction rates formulation}
We begin by considering a reaction system consisting of $n_s$ species and $n_r$ reactions, when a chemical equation described by the following:
\begin{equation}\label{eq:chem}
    \sum\limits_{i=1}^{n_s} \nu'_{ij} \mathcal{S}_i \rightleftharpoons \sum\limits_{i=1}^{n_s} \nu''_{ij} \mathcal{S}_i
\end{equation}
for $j = 1,2, ..., n_r$. In Eq. \ref{eq:chem}, $\nu'_{ij}$ and $\nu''_{ij}$ represents the reactant and product stoiciometric coefficients for respectively, for the $i-$th reaction and the $j-$th species, while $\mathcal{S}_j$ is a generic symbol representing a species. We also denote the the stoichiometric coefficient matrix, with components $\nu_{ij} = \nu'_{ij} - \nu''_{ij}$ as $\bm V \in \mathbb{R}^{n_s \times n_r}$. The forward rate coefficient, $k_f$, is based on the Arrhenius form
\begin{equation}
    k_f \left( T\right) = A \exp \left(-E/R_u T \right),
\end{equation}
where $A$ is the pre-exponential or frequency factor, $E$ is the activation energy, $R_u$ is the universal gas constant, and $T$ is the temperature. Based on these, the net rate-of-progress  variable for reaction $j$ can be computed as,
\begin{equation}\label{eq:rate_of_prog}
    q_j = k_{f,j} \prod_{i=1}^{n_s}\left[ \mathcal{S}_i\right]^{\nu'_{ij}} - k_{r,j} \prod_{i=1}^{n_s}\left[ \mathcal{S}_i\right]^{\nu''_{ij}}
\end{equation}
where $k_{r,j}$ is the reverse rate constant for reaction $j$. As opposed to defining its individual components as in Eq. \ref{eq:rate_of_prog}, we can define a column vector containing elements of the rate-of-progress variable as $\bm q = \left[q_1, q_2, ..., q_{n_r} \right]^T \in \mathbb{R}^{n_r \times 1}$. Finally, the vector of net reaction rates of the species is denoted $\bm \omega = \left[\omega_1, \omega_2, ..., \omega_{n_s} \right]^T \in \mathbb{R}^{n_s \times 1}$ and can be expressed as:
\begin{equation} \label{eq:net_reac_rates}
    \bm \omega = \bm{V} \bm{q}.
\end{equation}

\subsection{Quantized learning approach \label{subsec:qla}} \addvspace{10pt}

\begin{figure}[htbp]
\centering
\includegraphics[width=380pt]{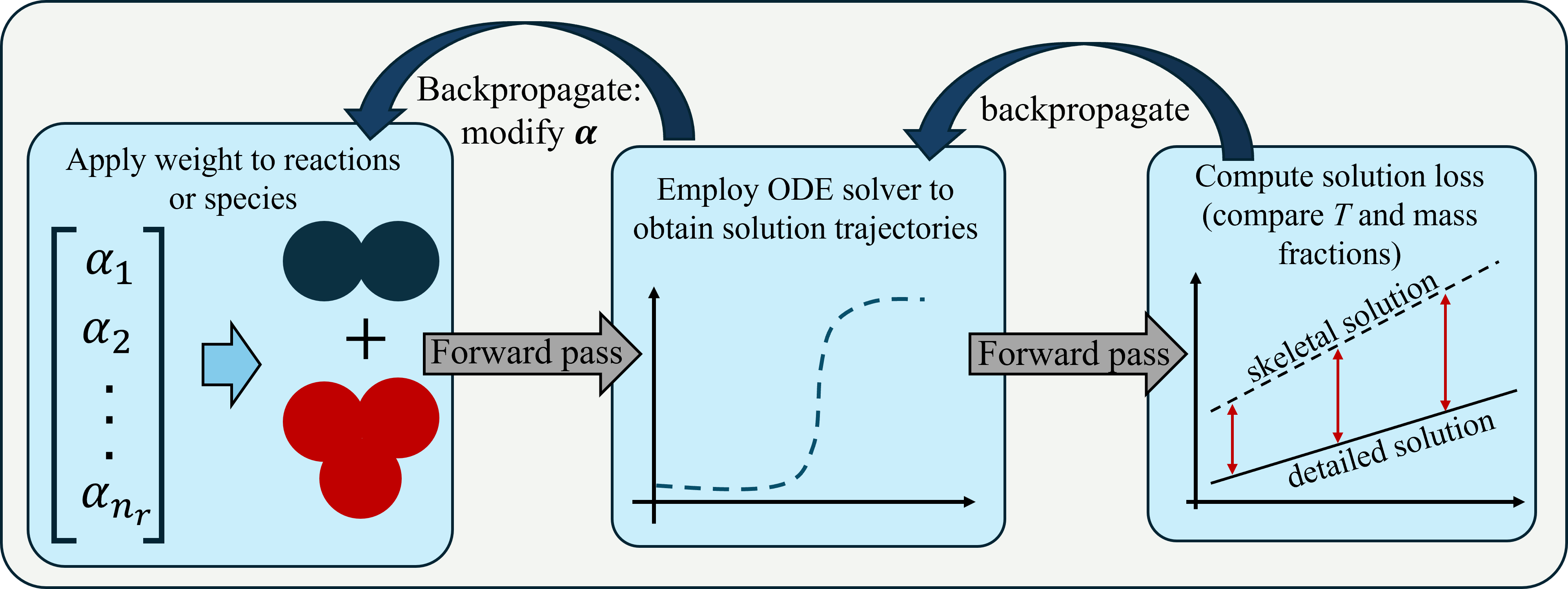}
\caption{\footnotesize Illustration of QSL approach, involving a forward pass where weighted source terms are employed to advance the solution in time, followed by a backpropagation through all the operations of the solver and chemical kinetics model to modify the weighting vector, $\bm \alpha$.}
\label{fig:num_reactions}
\end{figure}

The proposed skeletal reduction approach operates by minimizing the differences between the mass fraction profiles generated from a detailed chemical mechanism and a simplified version of this mechanism, with some relatively unimportant reactions or species removed. In this regard, the first step is to compute a solution to the system of ODEs that define the chemical system. In this study, skeletal reduction is performed based on data generated from zero-dimensional homogenous reactors at constant pressure. In this case, we have a coupled system of ODEs defined as:
\begin{equation}\label{eq:species}
    \frac{d \bm y}{dt} = \bm{f_Y} \left(p, T, \bm Y\right) = \frac{\bm m \circ \bm \omega}{\rho}
\end{equation}
where $\circ$ is the Hadamard product, $\bm y = \left[ y_1, y_2, ..., y_{n_s} \right]^T \in \mathbb{R}^{n_s \times 1}$ is the vector containing the species mass fractions, $\bm w = \left[m_1, m_2, ..., m_{n_s} \right]^T \in \mathbb{R}^{n_s \times 1}$ is a vector containing the species molecular weights, and $\rho$ is mixture density, computed using the ideal gas equation of state. The ODE governing the evolution of the mass fraction also couples with another ODE that determines the evolution of temperature:
\begin{equation}\label{eq:temp}
    \frac{d T}{dt} = - \frac{\bm{h}^T \bm{f_Y} } {c_p}
\end{equation}

where $\bm{h} = \left[ \Delta h_{f,1}^o, \Delta h_{f,2}^o, ..., \Delta h_{f,n_s}^o \right]^T \in \mathbb{R} ^ {n_s \times 1}$ is a vector containing the mass enthalpies of formation for the species. In Eq. \ref{eq:temp}, $c_p = \sum_{i = 1}^{n_s} c_{p,i} \left( T\right) y_i$, is the mixture specific heat at constant pressure on a mass basis, with $c_{p,i}$ being the specific heats at constant pressure of the components of the mixture, computed using the NASA polynomials \cite{mcbride2002nasa}. 
%

By solving Eqs. \ref{eq:species} and \ref{eq:temp}, the solution to the thermochemical vector at various locations in time is obtained and can be stored in a solution matrix.

As opposed to focusing on the reaction rates, the proposed QSL approach operates directly on the mass fractions obtained after integrating the skeletal and detailed mechanisms' mass fractions. The proposed QSL approach tries to minimize the difference between the thermochemical solution at various time instances for a zero-dimensional constant pressure autoignition problem. We denote the thermochemical vector at a given time instance as $\bm \psi = \left[T, y_1, y_2, ..., y_{n_s} \right]^T$, and the thermochemical matrix, composed of various realizations of the thermochemical vector, collected at "$n_o$" observations (i.e., at various time instances based on different initial conditions) as, $\bm \Psi = \left[\bm\psi_1, \bm\psi_2, ..., \bm\psi_{n_o} \right]^T$. In this case, the conventional mean squared error loss will be defined as:
\begin{equation}\label{eq:mse_loss}
    \mathcal{L}^{MSE} = \frac{1}{n_s n_o} \| \bm \Psi - { \bm \Psi}_r\|^2_F
\end{equation}
In Eq. \ref{eq:mse_loss}, $\bm \Psi$ is the solution matrix as obtained from the full chemical mechanism, while $\bm \Psi_r$ is the solution matrix as obtained from the skeletal mechanism. The primary goal remove unimportant reactions, while minimizing the loss function. In other words, reactions and species that minimally impact the deviation of the skeletal mechanism's species profiles from those obtained from the detailed mechanism are discarded, while those that significantly affect it are retained. 

To achieve this objective, the rate of reaction, as defined in \ref{eq:rate_of_prog} is modified using a weighting vector before this is used to compute the net reaction rates as shown in Eq. \ref{eq:net_reac_rates}. In the case of species elimination, we denote this as $\bm \alpha_s \in \mathbb{R}^{n_s \times 1}$, and in the case of eliminating reactions, we denote it as $\bm \alpha_s \in \mathbb{R}^{n_r \times 1}$. First, we consider the case where we seek to reduce the number of reactions, as this is the more straightforward case. In this case, the vector of net reaction rates is calculated as
\begin{equation}\label{eq:qsr-r}
    \bm \omega_r = \bm V \left( \bm q \circ \bm \alpha_r \right) 
\end{equation}
The ground truth solution used to calculate the loss in Eq. \ref{eq:mse_loss} is computed using this modified equation. By adjusting the values of the elements of $\alpha_r$, various reactions can be amplified or repressed, with a value of 1.0 signifying that the reaction is unmodified. In cases where the weight decays to zero, such reactions have been effectively eliminated altogether. Although the QSL-R adjusts the reaction weights, species are also removed once all the reactions they participate in are eliminated.

QSL can also be applied to the direct elimination of species, referred to as QSL-S in this paper. However, simply multiplying species by a weight vector is not feasible because reactions involving the removed species would lack the necessary reactants or products. One potential solution is to eliminate the entire reaction when any of its dependent species is removed. However, the proposed approach does not follow a binary learning scheme, where species/reactions are either present or absent. Rather, the weights are gradually decayed to 0 or increased to 1, making the abrupt removal of a reaction infeasible. Therefore, a different approach that is suitable for a gradient-based optimiation is followed. First, we compute an intermediate matrix, $\bm \Xi$ as
\begin{equation}\label{eq:xi}
    \bm \Xi = 1 + \bm{1} \bm \alpha_s^T - \mathrm{clip}\left( | \bm V |, 1.0 \right)
\end{equation}
In Eq. \ref{eq:xi}, $\bm 1_{n_r} = \left[ 1, 1, ... , 1 \right]^T \in \mathbb{Z}^{n_r \times 1}$, and $\mathrm{clip} \left( \cdot, 1.0 \right)$ is a function that clips all the entries of its argument matrix to a maximum of 1.0. Based on this, in the case of QSL-S, the net reactor rates are computed as
\begin{equation}
    \bm \omega = \bm V \left( \bm q \circ \bm \xi_m \right) 
\end{equation}
where  
\begin{equation}
    \bm \xi_m = \min\limits_{j} \left(\bm \Xi \right)
\end{equation}
with $\min\limits_{j}$ representing the elementwise minimum across the columns of $\bm \Xi$ for each row.

\subsection{Optimization process\label{subsec:opt}} \addvspace{10pt}
The first optimization goal is to optimize for the entries of $\bm{\alpha}_r$ or $\bm{\alpha}_s$ such that the errors between the profiles of the thermochemical species obtained from the skeletal and full mechanism, as computed in $\bm {\Psi}$, is minimized. The second goal is to achieve sparsity in $\bm \alpha$, i.e., we want some of the entries to decay to zero. In both cases, $\bm \alpha$ scales the mass fraction of that species or the reaction rates, thus controlling their contributions to the final solution. By decaying some of the elements to zero, species and reactions are essentially being eliminated. Finally, the third goal is to constrain the values of $\alpha$ within the range [0, 1], as it is not meaningful to scale species or reaction rates by negative values or values greater than unity.

\begin{figure}[h!]
\centering
\includegraphics[width=220pt]{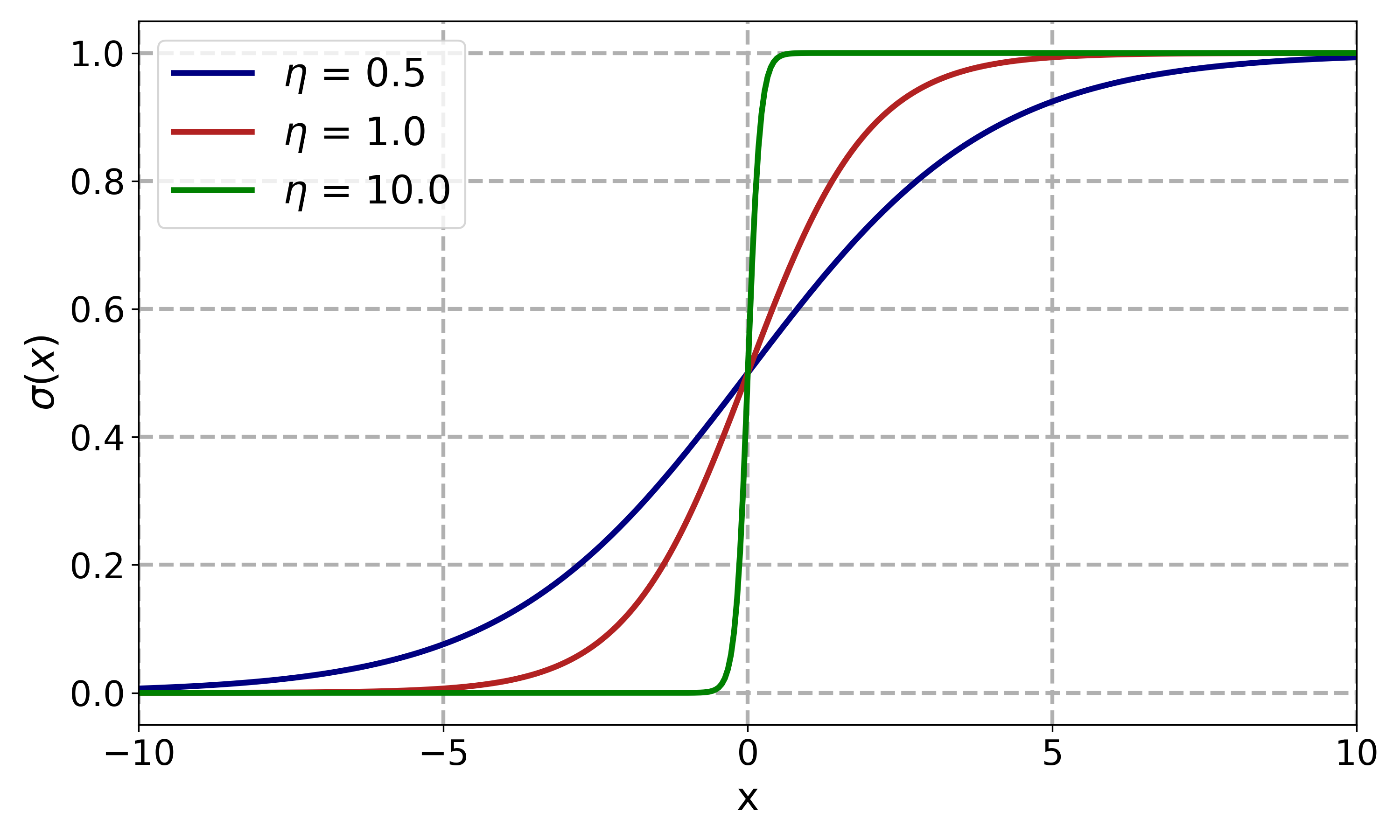}
\caption{\footnotesize Sigmoid activation used to constrain weights to the interval [0, 1].}
\label{fig:sigmoid}
\end{figure}

To prevent the values of $\bm \alpha$ from leaving the bounds of zero to 1, it is not directly modified during training. Rather, we adjust another surrogate variable, $\bm {\hat{\alpha}}$ instead, related to $\bm \alpha$ by a modified sigmoid function:
\begin{equation}
    \bm \alpha =  \frac{1}{1 + e^{-\eta \bm {\hat{\alpha}}}}
\end{equation}
$\eta$ controls the slope of the sigmoid function, as shown in Fig. \ref{fig:sigmoid}. Lower values of $\eta$ result in a gentler slope, while higher values produce a steeper slope. The standard sigmoid function corresponds to $\eta = 1.0$. The values of $\bm{\alpha}$ range from 0 to 1, whereas the values of $\bm{\hat{\alpha}}$ range from $-\infty$ to $+\infty$.

We devise a loss function that incorporates these three optimization objectives discussed above:
\begin{equation}\label{eq:loss}
    \mathcal{L} = \frac{\| \bm \Psi - \bm \Psi_r\|^2_F}{n_s n_r}  + \kappa \frac{ \| \bm{ \hat{\alpha} }\|_1}{n}
\end{equation}
where $n = n_s$ and $\bm \alpha = \bm \alpha_s$ for the QSL-S variant where species are directly eliminated, and $n = n_r$ and $\bm \alpha = \bm \alpha_r$ for QSL-R, where reactions are being eliminated. The second term on the right-hand side of Eq. \ref{eq:loss} is a $l_1$-regularization term that encourages the optimizer to promote sparsity in $\bm{ \hat{\alpha} }$ by decaying some of its values to zero, thus eliminating some species and reactions. Higher values of $\kappa$ promote sparser representations, and vice-versa

During each iteration, the values of $\bm \hat{\alpha}$ are progressively adjusted in the direction of steepest descent on the loss landscape. The value of $\hat{\bm \alpha}$ as iteration $k+1$ are computed based on its values at iteration $k$ using the Adam optimizer \cite{kingma2014adam}. During training, the value of $\eta$ is gradually increased, inspired by the approach employed in a previous study \cite{yang2019quantization}. This has the effect of forcing values in $\bm \alpha$ to jump in a more discontinuous fashion as the training progresses (as seen in Fig. \ref{fig:sigmoid}, pushing the values of $\bm \alpha$ to either settle close to 0 or 1, the extremities of the sigmoid function.

\section{Results and discussion}
In this section, we present the results of applying the QLR algorithm to the skeletal reduction of the GRI-Mech 3.0 mechanism for methane-air combustion. To generate training data, the detailed mechanism was used to simulate a constant-pressure reactor under initial conditions with temperatures ranging from 950 to 1200 K, equivalence ratios from 0.5 to 1.5, and pressures ranging from 1 atm to 30 atm. For all these cases, temperature and mass fraction profiles were stored as a function of time.

A parallelized batch training algorithm was employed, where approximately 11\% of the full dataset (16 solution profiles out of 144) was randomly selected at each iteration to compute the gradients of the loss and update $\hat{\bm{\alpha}}$. To enable end-to-end automatic differentiation, the entire workflow -- computing physical properties, reaction rates, and the subsequent temporal integration using a backward differentiation formula -- was implemented in JAX. For all cases, 3 trials were performed, and the case with the best MSE was retrained. For 7,000 iterations, the value of $\eta$ was ramped up, where the value of $\eta$ at iteration $k$ is given by $\eta^{(k)} = \gamma \eta^{(k-1)}$, where $\gamma$, the ramp factor, is chosen to be a constant value such that $\eta \approx 100.0$ at $k = 7,000$. After iteration 7,000, the value of $\eta$ is frozen for an additional 1,000 iterations where the weights of the lowest achieved MSE value are saved before the training process terminated. A constant learning rate of $5 \times 10^{-3}$ was used to adjust $\hat{\bm{\alpha}}$ in conjunction with the Adam optimizer.

\subsection{Direct elimination of reactions\label{subsec:reactions_elimination}} \addvspace{10pt}


\begin{figure}[htbp]
\centering
\includegraphics[width=180pt]{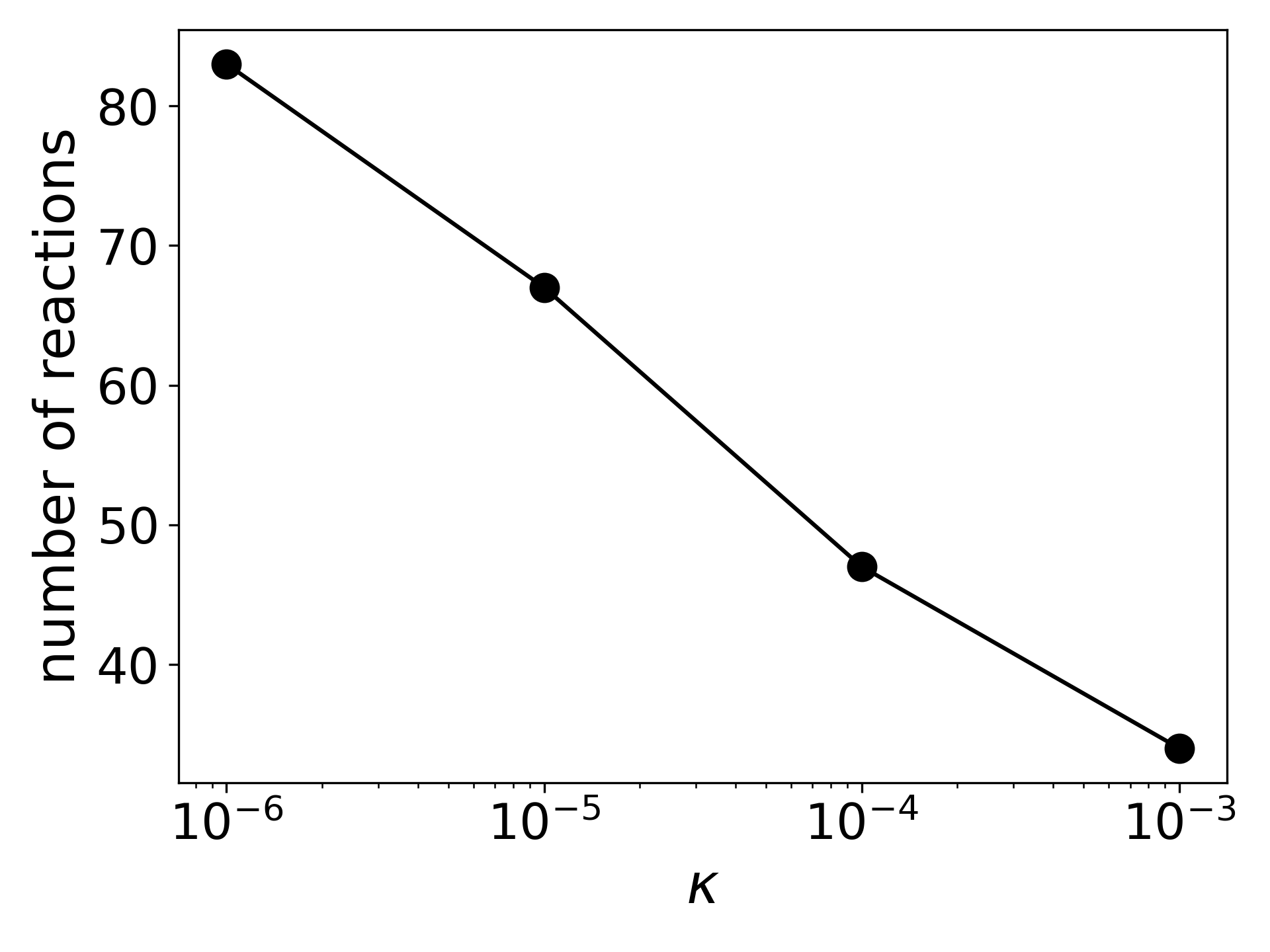}
\caption{\footnotesize Number of reactions as a function of the level of regulariation.}
\label{fig:num_reactions}
\end{figure}

As a first test, we present results using the QSR-R variant, which involves computing the reaction rates using Eq.~\ref{eq:qsr-r}, where the goal is to directly eliminate less important reactions. Species are eliminated indirectly: a species is removed once all reactions involving that species have been eliminated. In this case, instead of using the full set of species to compute the loss function as shown in Eq.~\ref{eq:loss}, the following scalars were selected as targets for learning: Temperature, \(y_{CH_4}\), \(y_{CO_2}\), \(y_{O_2}\), and \(y_{H_2O}\). The value of \(\kappa\), the parameter controlling the level of regularization applied to the loss, was varied logarithmically from \(10^{-6}\) to \(10^{-3}\). For brevity, the following shorthand is sometimes used throughout the remainder of this manuscript: \(\kappa = 10^{-3}\) is denoted as \(\kappa3\), \(\kappa = 10^{-4}\) as \(\kappa4\), \(\kappa = 10^{-5}\) as \(\kappa5\), and \(\kappa = 10^{-6}\) as \(\kappa6\).

Higher values of \(\kappa\), based on the framing of the loss function in Eq.~\ref{eq:loss}, are expected to produce sparser \(\bm{\alpha}\) and, consequently, more aggressive eliminations, leading to fewer reactions. In such cases, the algorithm prioritizes smaller mechanism sizes over fidelity to the full mechanism.

\begin{table}[h!] \footnotesize

\caption{Number of species and reactions for various levels of regularization using QSL-R.}
\centerline{\begin{tabular}{c c c c c}
\hline 
\multicolumn{1}{c}{} & \multicolumn{4}{c}{$\kappa$} \\
\cmidrule(rl){2-5}
                      & $10^{-3}$ & $10^{-4}$ & $10^{-5}$ & $10^{-6}$     \\
\hline
species                              & 19     & 26 &  27  & 30\\
reactions                      & 34 & 47  &   67  &  83\\
\hline 
\end{tabular}}

\label{tab:qsrr}
\end{table}

Table~\ref{tab:qsrr} summarizes the number of species and reactions retained for various levels of regularization. As expected, higher values of \(\kappa\) induce greater sparsity in \(\bm{\alpha}_r\), the weighting vector, resulting in fewer retained reactions and, as a consequence, fewer retained species. The retained species and reactions range from 30 species and 84 reactions (for \(\kappa = 10^{-6}\)) to just 19 species and 34 reactions (for \(\kappa = 10^{-3}\)). As shown in Fig.~\ref{fig:num_reactions}, this corresponds to an approximately exponentially decaying relationship between \(\kappa\) and the number of retained reactions.

\begin{figure}[htbp]
\centering
\includegraphics[width=180pt]{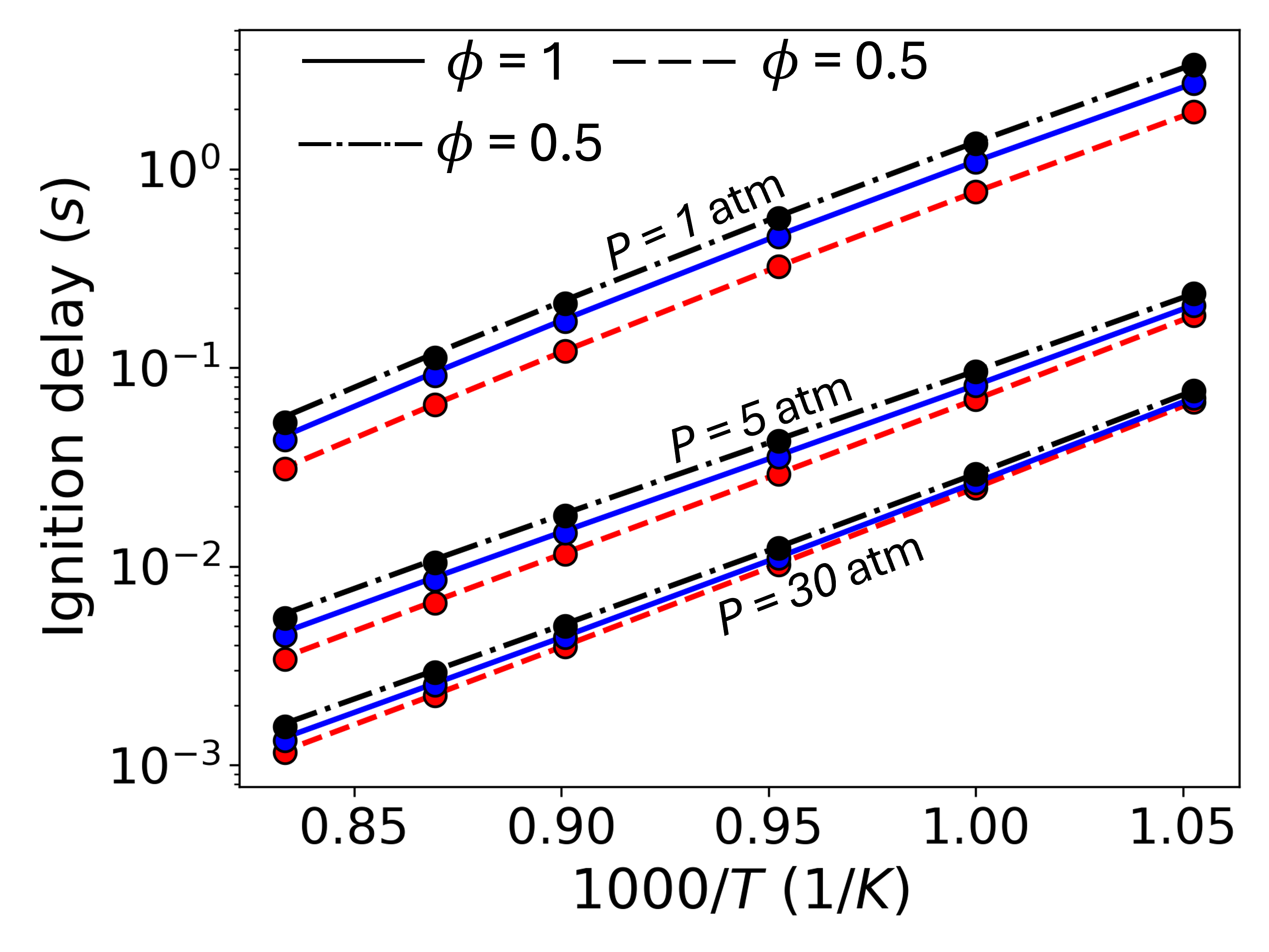}
\caption{\footnotesize Ignition delay as a function of 1000/T for various levels of equivalence ratio and pressure for $\kappa = 10^{-4}$.}
\label{fig:idg_reactions}
\end{figure}

Fig.~\ref{fig:idg_reactions} shows the ignition delay obtained from the \(\kappa3\) case compared to the ignition delay from the full mechanism as a function of \(1000/T\) for various pressure levels. Only the \(\kappa4\) case is shown -- higher levels of \(\kappa\) produced significant errors, while lower levels appeared visually similar. As seen in the figure, the skeletal mechanism generated by QSL-R produces ignition delay results that are in excellent agreement with the detailed mechanism. The maximum and mean absolute errors in ignition delays for the \(\kappa4\) case are 2.4\% and 0.23\%, respectively.

The \(\kappa3\) case, with its lower number of retained species and reactions, as expected, produces higher errors, with maximum and mean errors of 50\% and 4.15\%, respectively. For \(\kappa = 10^{-5}\), the corresponding values are 5.7\% and 0.45\%, while for \(\kappa = 10^{-6}\), they are 5.8\% and 0.5\%. The trends in ignition delay errors beyond \(\kappa4\) are somewhat counterintuitive, as one might expect the ignition delay errors to monotonically increase as fewer species and reactions are retained. However, it should be noted that the QSL learning approach, as described by the loss function in Eq.~\ref{eq:loss}, is not directly optimized for ignition delay but rather for the mean squared error (MSE) of normalized species concentrations. Observing the MSE of species concentrations reveals a more consistent trend, with MSE values of \(4.37 \times 10^{-2}\), \(5.58 \times 10^{-3}\), \(1.25 \times 10^{-3}\), and \(8.17 \times 10^{-4}\) for the \(\kappa3\), \(\kappa4\), \(\kappa5\), and \(\kappa6\) cases, respectively.

\begin{figure*}[t]
\centering
\includegraphics[width=320pt]{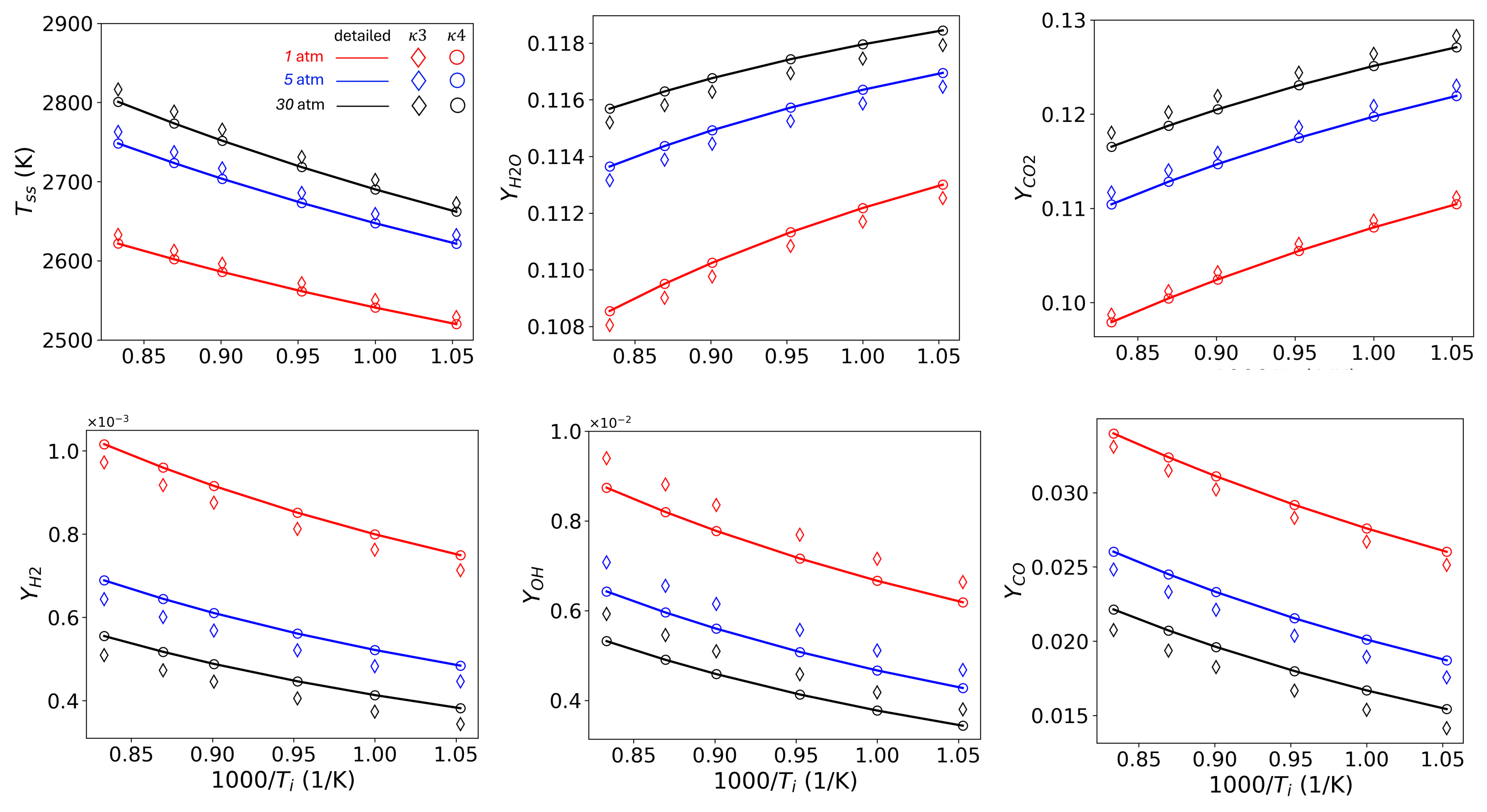}
\caption{\footnotesize Steady-state concentrations of thermochemical scalars.}
\label{fig:ssA}
\end{figure*}

Fig.~\ref{fig:ssA} shows the steady-state concentrations of various species as functions of \(1000/T\) for different pressure levels. As seen in the figure, the \(\kappa3\) case with 19 species and 34 reactions captures trends accurately across all conditions for the scalars considered. However, quantitatively significant errors can be observed, particularly for OH and H$_2$, where the mean percentage errors are 11\% and 9\%, respectively. These errors are significantly reduced in the \(\kappa4\) case with 26 species and 47 reactions, where much better agreement is observed across all thermochemical scalars. The worst case in this scenario is an error of 0.016\%, which occurs for OH. 

\subsection{Direct elimination of species \label{subsec:species_elimination}} \addvspace{10pt}

\subsubsection{\ce{CH4} mechanism reduction}
\begin{figure*}[htbp]
\centering
\includegraphics[width=280pt]{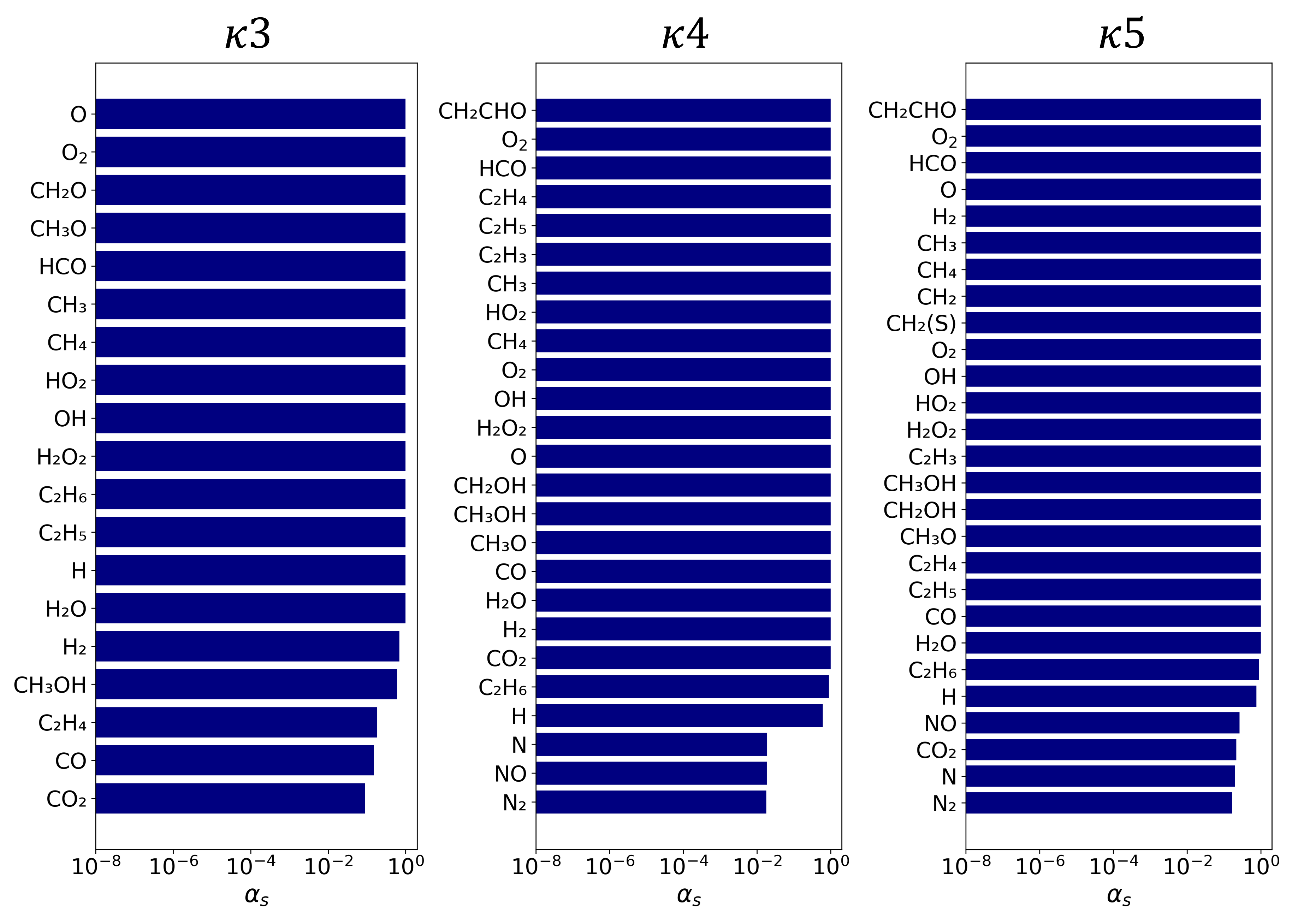}
\caption{\footnotesize Weights of retained species obtained with various levels of regularization.}
\label{fig:weights}
\end{figure*}

This section presents the results obtained using the QSL-S variant, which targets a reduction of the species. As described in the formulation laid out in section \ref{subsec:opt}, reactions are also eliminated on the values of the weight vector, $\bm {\alpha}$. Apart from the differences with the QSL-R already described in that section, the setup, including the range of conditions used to train, learning rates, and other model parameters remain identical. Fig. \ref{fig:weights} shows the final values of the weights assigned to each species based on the elements of vector $\bm{\alpha}$ after training is complete. In the figure, values below $10^{-6}$ are not shown -- the weights of species below this threshold are deemed negligible, and thus, rounded down to zero. On the other hand, species with weights greater than this threshold are retained.

\begin{table}[h!] \footnotesize
\caption{Number of species and reactions for various levels of regularization using QSL-S.}
\centerline{\begin{tabular}{c c c c c}
\hline 
\multicolumn{1}{c}{} & \multicolumn{4}{c}{$\kappa$} \\
\cmidrule(rl){2-5}
                      & $10^{-3}$ & $10^{-4}$ & $10^{-5}$ & $10^{-6}$     \\
\hline
number of species                              & 21     & 26 &  28  & 28\\
number of reactions                      & 86 & 113  &   145  &  145\\
\hline 
\end{tabular}}

\label{tab:mechsize}
\end{table}

Table \ref{tab:mechsize} shows how the number of species and reactions vary as the value of $\kappa$ is adjusted. At $\kappa = 10^{-3}$, only 21 species and 86 reactions are retained, while at the weakest regularization ($\kappa = 10^{-6}$), the number of species and reactions are 28 and 145, respectively. All species and reactions related to NO chemistry are eliminated for all but the weakest regularization ($\kappa = 10^{-6}$). For this case, the number of species and reactions retained for the $\kappa 5$ and $\kappa 6$ levels of regularization are identical, possibly due to the randomness of training, or perhaps, indicating the diminishing role the regularization term plays in eliminating species below $\kappa = 10^{-5}$. Compared to the full mechanism, the case with \(\kappa = 10^{-6}\) primarily eliminates nitrogen-containing species (e.g., nitrogen oxides, HCN, and its derivatives), pure carbon species and carbon radicals, some \(C_2\) hydrocarbons, oxygenated hydrocarbons, and argon. When \(\kappa\) is increased to \(10^{-4}\), methylenes (\(CH_2\) and \(CH_2(S)\)) are further eliminated. Additional eliminations associated with the $\kappa 3$ case include \(CH_2CHO\), \(CH_2OH\), \(C_2H_3\), \(N\), \(NO\), and \(N_2\), although \(N_2\) is manually added back after the training process.

\begin{figure}[htbp]
\centering
\includegraphics[width=180pt]{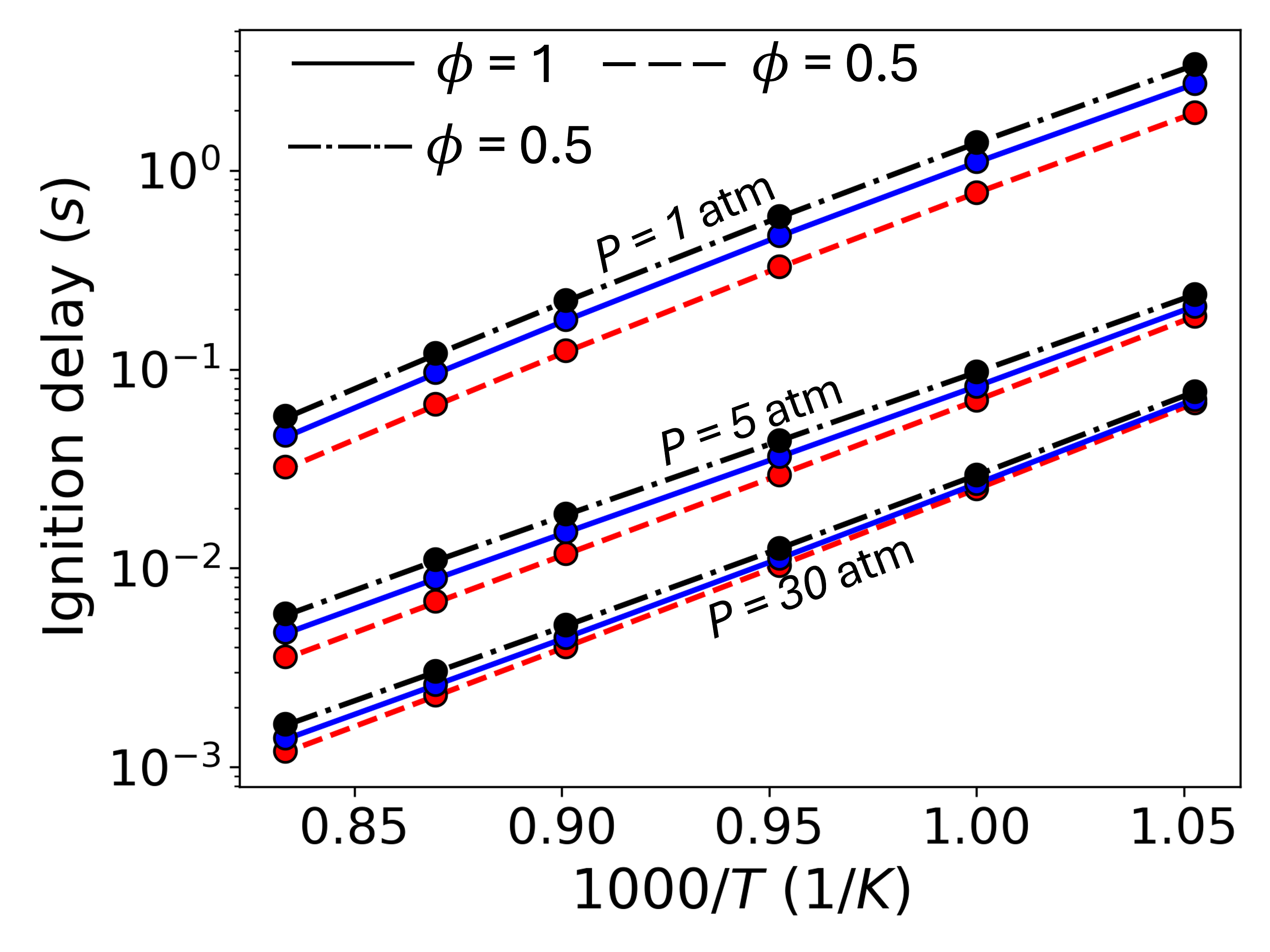}
\caption{\footnotesize Ignition delay as a fuunction of 1000/T for various levels of equivalence ratio and pressure for $\kappa = 1\times 10^{-3}$.}
\label{fig:idg_species}
\end{figure}


The ignition delays as a function of \(1000/T\) for various levels of \(\phi\) and pressures are shown in Fig.~\ref{fig:idg_species}. Only the case with the strongest regularization (\(\kappa3\)), representing the worst-case scenario in terms of accuracy, is shown in this figure, as lower levels of regularization lead to even better results. Overall, the skeletal mechanism captures the qualitative trends and quantitative values of ignition delays observed in the full mechanism, correctly displaying the inverse relationship with temperature and pressure. Other values of \(\kappa\) are not shown visually—since the \(\kappa = 10^{-3}\) case shows good visual agreement, the other cases can be expected to perform similarly well. In general, the maximum errors in ignition delays are 3.3\%, 0.24\%, 0.31\%, and 0.31\% for the \(\kappa = 10^{-3}\), \(10^{-4}\), \(10^{-5}\), and \(10^{-6}\) cases, respectively. Similar to the QSL-R case, the \(\kappa5\) case produces better species profile errors compared to \(\kappa4\), but results in slightly worse ignition delay times.

\begin{figure*}[htbp]
\centering
\includegraphics[width=280pt]{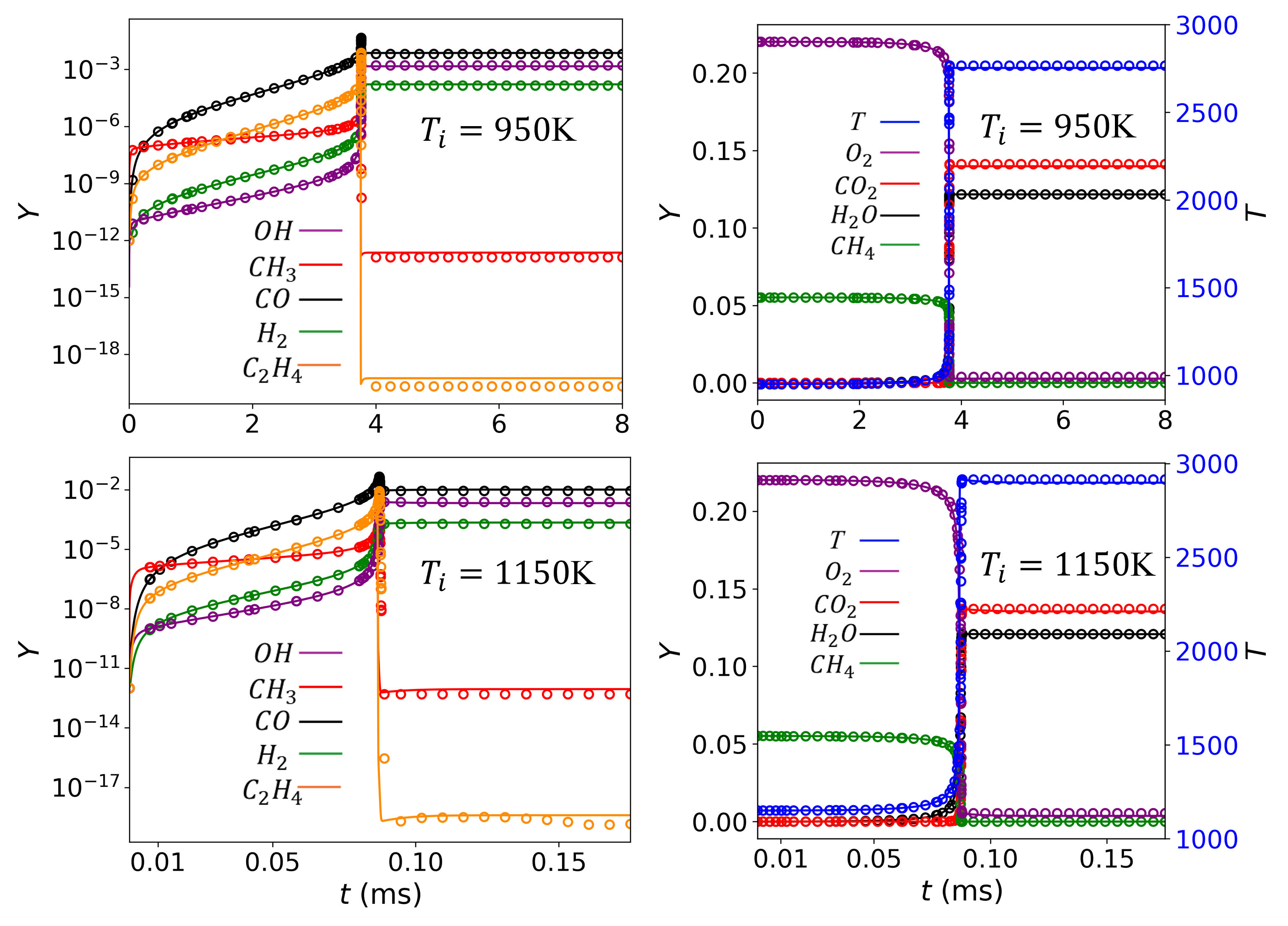}
\caption{\footnotesize Temporal profiles of profiles of temperature, major reactants, and products (left), and selected intermediates species for initial temperatures of 950 K (top) and 1150 K (bottom). The lines represent the solution obtained using the detailed GRI-Mech 3.0 mechanism, while the symbols represent the solution from the $\kappa 3$ skeletal mechanism.}
\label{fig:0D3}
\end{figure*}

\begin{figure*}[htbp]
\centering
\includegraphics[width=350pt]{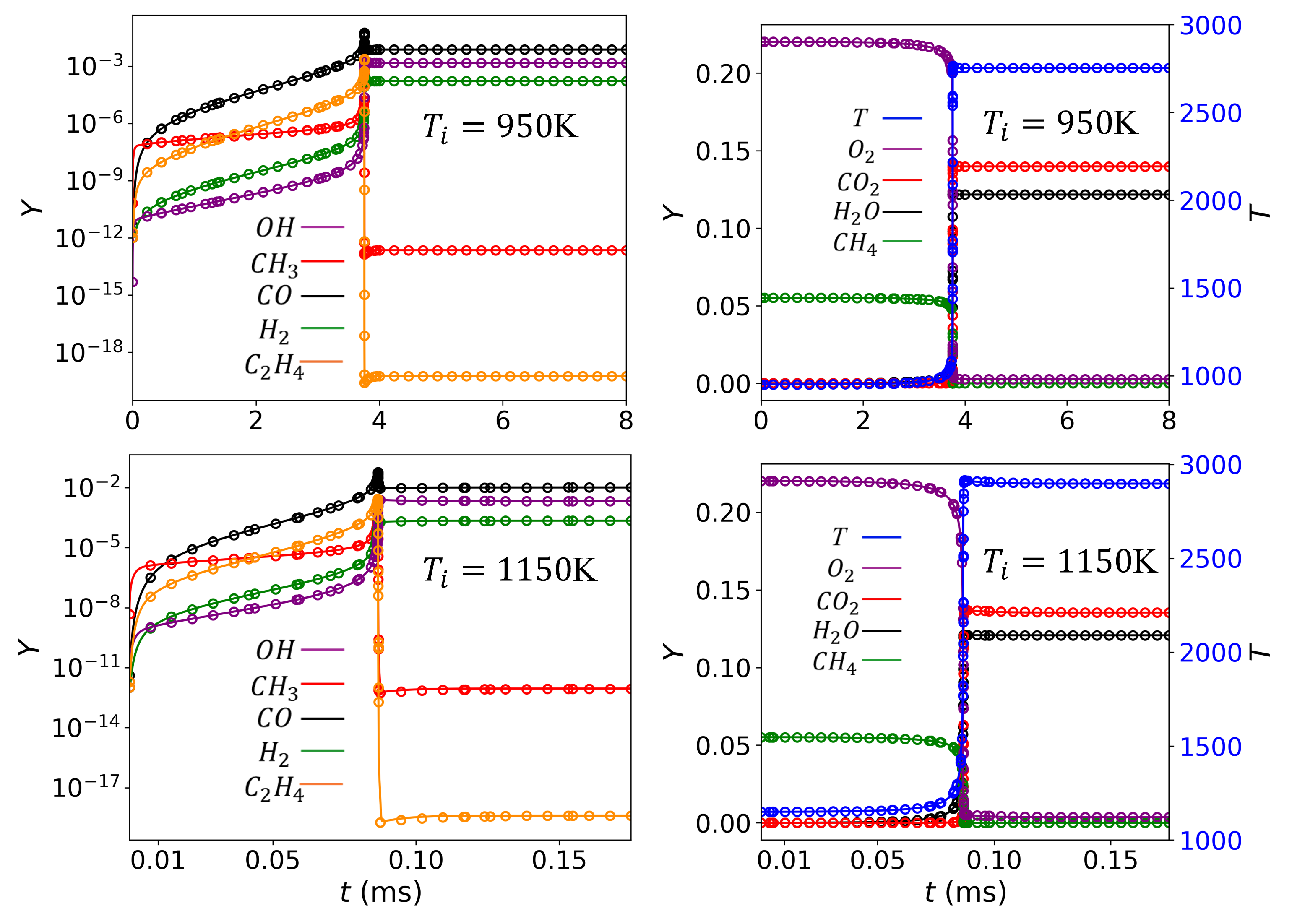}
\caption{\footnotesize Temporal profiles of profiles of temperature, major reactants, and products (left), and selected intermediates species for initial temperatures of 950 K (top) and 1150 K (bottom). The lines represent the solution obtained using the detailed GRI-Mech 3.0 mechanism, while the symbols represent the solution from the $\kappa 4$ skeletal mechanism.}
\label{fig:0D4}
\end{figure*}

The profiles of selected species under specific conditions are also shown in Figs.~\ref{fig:0D3} and \ref{fig:0D4} for the \(\kappa3\) and \(\kappa4\) cases, respectively, representing mechanisms with 21 species and 26 species. Both figures illustrate temperature, major products and reactants, and some intermediate species for initial temperatures of \(T_i = 950~K\) and \(T_i = 1150~K\). Fig.~\ref{fig:0D4} shows that the skeletal mechanism generated with the \(\kappa4\) case achieves excellent agreements with the detailed mechanism for temperature, reactants, products, and intermediate species. For the \(\kappa3\) case, some discrepancies can be observed, particularly in the steady-state behavior of \(C_2H_4\) and minor underprediction or overprediction of certain scalars (e.g., \(CO_2\), temperature, \(CH_3\)).

\subsubsection{Direct elimination of species with NO chemistry \label{subsec:NO_elimination}} 

The QSL algorithm can be tailored to a desired set of thermochemical scalars by modifying the list of species used to compute the MSE in Eq.~\ref{eq:loss}. For the cases presented in Sections~\ref{subsec:reactions_elimination} and~\ref{subsec:species_elimination}, the following scalars were used: Temperature, \(y_{CH_4}\), \(y_{CO_2}\), \(y_{O_2}\), and \(y_{H_2O}\). Here, we modify the list of scalars to include \(y_{NO}\) and \(y_{NO_2}\), while retaining the initial list of scalars. The QSL-S variant, which targets the elimination of species (as opposed to reactions), is used for this analysis. Apart from the change in training variables, all other training parameters remain consistent with those in Section~\ref{subsec:species_elimination}, including learning rates, ramp-up factors, and the number of training iterations. However, instead of comparing various levels of regularization, only \(\kappa = 10^{-4}\) is considered here.

\begin{figure}[htbp]
\centering
\includegraphics[width=380pt]{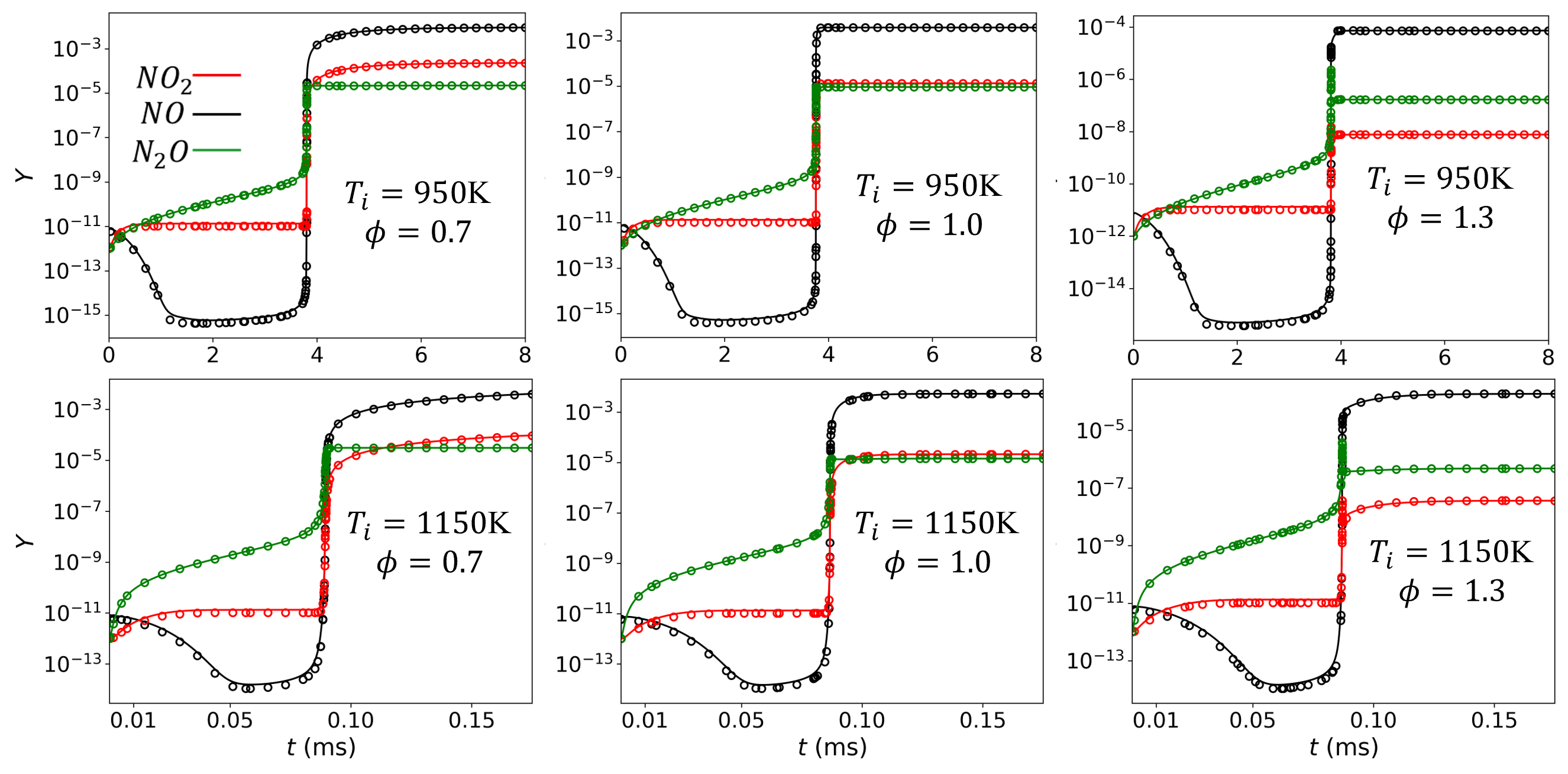}
\caption{\footnotesize Temporal profiles of profiles of $NO$, $NO_2$ and $N_2O$ for various initial temperatures and equivalence ratios. The lines represent the solution obtained using the detailed GRI-Mech 3.0 mechanism, while the symbols represent the solution from the skeletal mechanism.}
\label{fig:NO_profiles}
\end{figure}

Fig.~\ref{fig:NO_profiles} shows the temporal profiles obtained from a zero-dimensional constant-pressure reactor, comparing results from the detailed mechanism and the skeletal mechanism. The pressure is fixed at 1 atm for all cases, with three equivalence ratio levels (\(\phi\)) and two initial temperature levels considered. The results indicate that, particularly at low initial temperatures and lean conditions, the model accurately captures the more gradual approach to steady state for \(y_{NO}\) and \(y_{NO_2}\).

We also compute the relative errors for various temperatures and equivalence ratios using the formula 
\[
\text{Relative Error} = \frac{\left| \log(y_{QSL}) - \log(y_{\text{detailed}}) \right|}{\log(y_{\text{detailed}})},
\]
with the results shown in Table~\ref{tab:NO_errors}. These results demonstrate excellent agreement with the detailed mechanism, with the worst-case error occurring at \(1150~K\) and \(\phi = 0.7\) for $NO$ and $NO_2$, while for $N_2O$, the worst error occurs at leaner and low initial temperature conditions. 


\begin{table}[h!] \footnotesize
\caption{Relative errors of species mass fractions obtained using skeletal mechanism.}
\centerline{\begin{tabular}{c | c c c}
\hline 
\multicolumn{1}{c}{} & \multicolumn{3}{c}{$T_i = 950K$} \\
\cmidrule(rl){2-4}
                      $\phi$& $NO$ & $NO_2$ & $N_2O$    \\
\hline
0.7            & $5.8 \times 10^{-3}$ & $5.2 \times 10^{-3}$ &  $6.3 \times 10^{-5}$\\
1.0            & $4.5 \times 10^{-3}$ & $4.5 \times 10^{-3}$ &  $1.5 \times 10^{-5}$\\
1.3            & $4.9 \times 10^{-3}$ & $4.9 \times 10^{-3}$ &  $1.6 \times 10^{-4}$\\
\hline 
\multicolumn{1}{c}{} & \multicolumn{3}{c}{$T_i = 1150K$} \\
\cmidrule(rl){2-4}
                      $\phi$& $NO$ & $NO_2$ & $N_2O$    \\
\hline
0.7            & $9.5 \times 10^{-3}$ & $6.7 \times 10^{-3}$ &  $5.9 \times 10^{-5}$\\
1.0            & $7.9 \times 10^{-3}$ & $5.6 \times 10^{-3}$ &  $1.3 \times 10^{-4}$\\
1.3            & $7.9 \times 10^{-3}$ & $5.7 \times 10^{-3}$ &  $4.4 \times 10^{-4}$\\
\hline 
\end{tabular}}

\label{tab:NO_errors}
\end{table}

\subsubsection{USC Mechanism II Reduction \label{subsec:usc_elimination}} 

To further validate the QSL-S approach, we apply it to the USC Mech II mechanism, which consists of 111 species and 784 reactions, with the goal of extracting a skeletal mechanism for ethylene (\ce{C2H4}). The data used to train this reduced mechanism is generated using constant-pressure reactor simulations at initial temperatures ranging from 700–1450~K, pressures from 1–60~atm, and equivalence ratios between 0.5 and 2.0.
In this case, we vary the regularization parameter \(\kappa\), testing three values: \(\kappa = 10^{-3}\) ($\kappa 3$), \(\kappa = 10^{-4}\) ($\kappa 4$), and \(\kappa = 10^{-5}\) ($\kappa 5$). We also compare the resulting skeletal mechanisms against those generated using the Directed Relation Graph with Error Propagation (DRGEP) \cite{pepiot2008efficient} method.

\begin{figure}[htbp]
    \centering
    \includegraphics[width=380pt]{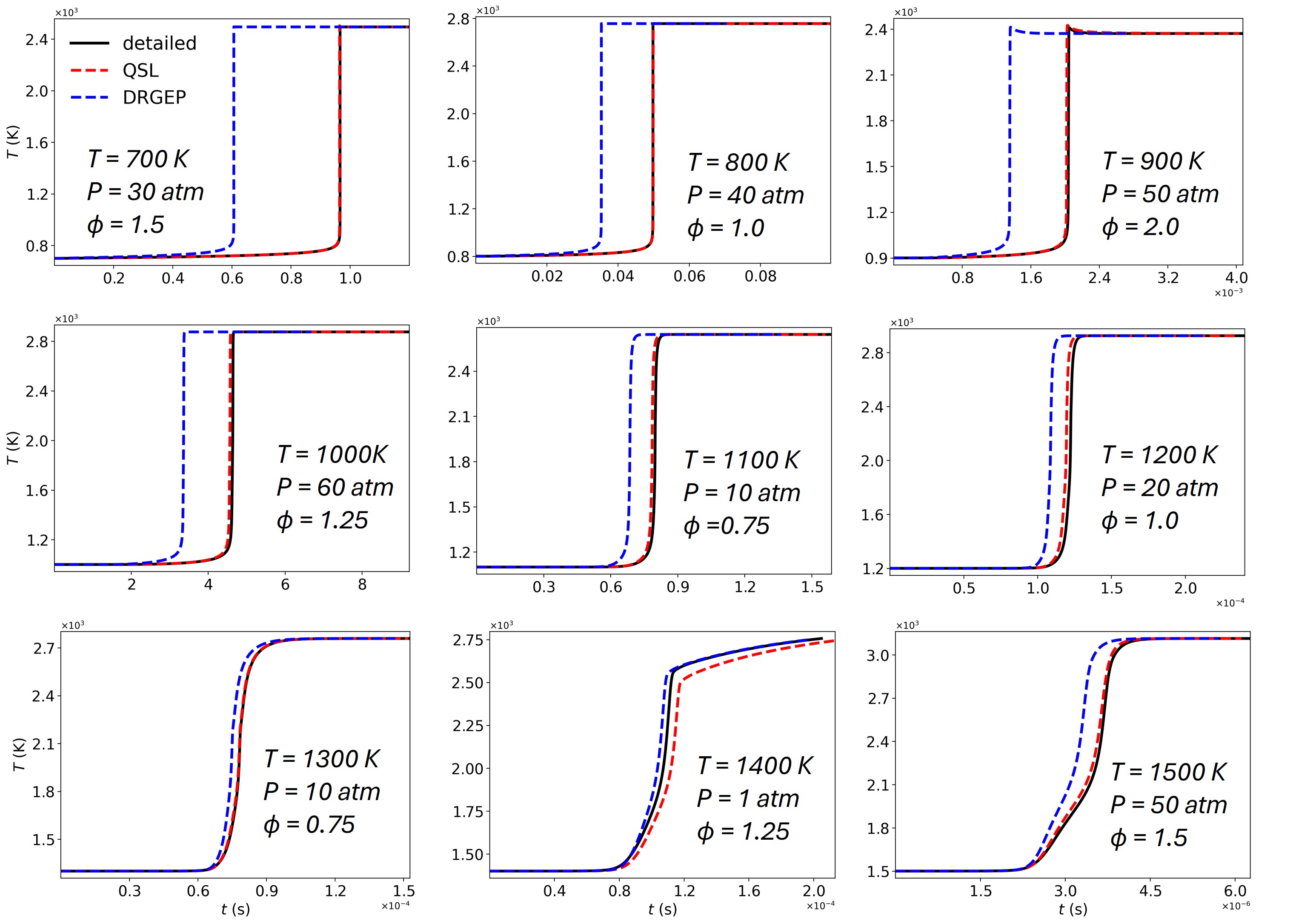}
    \caption{\footnotesize Temporal temperature profiles under various initial conditions (temperature, pressure, equivalence ratio), comparing the detailed mechanism (black solid line), a QSL-reduced mechanism with 29 species and 150 reactions (red dashed line), and a DRGEP-derived mechanism with 43 species and 275 reactions (blue dashed line).}
    \label{fig:usc_skeletal}
\end{figure}

For the DRGEP approach, we set a maximum allowable ignition delay error of 50\% as the reduction criterion. Based on this, we obtain a skeletal mechanism consisting of 43 species and 275 reactions. In contrast, the QSL-generated mechanisms for the $\kappa3$, $\kappa4$, and $\kappa5$ cases contain 29, 37, and 55 species, respectively, and 150, 236, and 371 reactions, respectively.

Figure~\ref{fig:usc_skeletal} shows temperature profiles from selected initial conditions within the training range, comparing predictions from the detailed mechanism, the DRGEP-derived mechanism, and the smallest QSL-reduced mechanism ($\kappa3$). Despite having significantly fewer species and reactions than the DRGEP counterpart, the QSL-reduced mechanism demonstrates significantly better predictive performance, particularly in terms of ignition delay times. Overall, the DRGEP-derived skeletal mechanism tends to systematically ignite earlier than the detailed mechanism, with discrepancies becoming more pronounced at lower temperatures and pressures.

\begin{table}[htbp]
\centering
\caption{\footnotesize Comparison of ignition delay errors and mechanism sizes for QSL and DRGEP approaches.}
\label{tab:mape_comparison}
\footnotesize
\setlength{\tabcolsep}{6pt}
\renewcommand{\arraystretch}{1.15}
\begin{tabular}{lcccc}
\toprule
\textbf{Mechanism} & \textbf{Species} & \textbf{Reactions} & \textbf{MAPE (\%)} & \textbf{Max APE (\%)} \\
\midrule
QSL (\(\kappa3\)) & 29 & 150 & 1.75 & 9.41 \\
QSL (\(\kappa4\)) & 37 & 236 & 0.69 & 2.69 \\
QSL (\(\kappa5\)) & 55 & 371 & 0.26 & 2.23 \\
DRGEP             & 43 & 275 & 17.20 & 39.35 \\
\bottomrule
\end{tabular}
\end{table}

Table~\ref{tab:mape_comparison} summarizes the Mean Absolute Percentage Error (MAPE) and Maximum Absolute Percentage Error (Max APE) in ignition delay obtained from the different reduction approaches. As expected, the QSL-derived skeletal mechanisms become progressively more accurate as the regularization parameter~\(\kappa\) decreases, indicating that the learning algorithm prioritizes fidelity over compactness. In all cases, the QSL mechanisms outperform the DRGEP approach, achieving significantly lower average and maximum ignition delay errors.

\subsection{Conclusion}
In this study, a Quantized Skeletal Learning (QSL) approach for generating skeletal mechanisms from detailed chemical mechanisms was introduced. This approach formulates mechanism reduction as a learning problem, where a weighting vector combined with a sigmoid activation function is optimized through a regularized, gradient-descent-based algorithm. We employ differentiable programming principles to solve the mechanism in a forward pass, before backpropagating through all the operations of the solver to eliminate less important species or reactions. Two variants of the QSL approach were presented: one focusing on reaction elimination (QSL-R) and the other on species elimination (QSL-S). The method was applied to methane chemistry reduction based on GRI Mech 3.0, first targeting key species without NO chemistry, and subsequently including NO chemistry, showing good agreement for \(\ce{NO}\), \(\ce{NO2}\), and \(\ce{N2O}\). Finally, the QSL-S approach was used to generate a skeletal mechanism for ethylene starting from the detailed USC Mech II mechanism, demonstrating superior accuracy compared to the DRGEP approach. Overall, the results show that the proposed QSL approach provides a promising data-driven framework for generating reliably accurate skeletal mechanisms, as the learning process is inherently coupled with time integration of the chemical system.

The limitations of the proposed method are as follows: although QSL was shown to generate more parsimonious and accurate mechanisms compared to DRGEP, it incurs higher computational costs, requiring \(\mathcal{O}(\text{hours})\) to a day on a single-core compute node. Therefore, in its current form, it is not well-suited for very large mechanisms containing several hundred species, since the method requires repeated integration and backpropagation through the solver's operations. In such cases, two practical strategies can be adopted: first, DRGEP may be used as a preliminary reduction step before a secondary QSL-based refinement; alternatively, the learning horizon can be limited to a fixed number of integration steps rather than spanning the entire simulation time. Strategies for limited GPU acceleration can also be explored. Future work will focus on incoporate these extensions and performing validation tests on larger mechanisms and more complex hydrocarbon fuels. 

\acknowledgement{Acknowledgments} \addvspace{10pt}
Portions of this research were conducted with high-performance computing resources provided by Louisiana State University (http://www.hpc.lsu.edu)
\end{spacing}

\newpage



 \footnotesize
 \baselineskip 9pt

\begin{spacing}{1.2}

\begin{thebibliography}{10}
\expandafter\ifx\csname url\endcsname\relax
  \def\url#1{\texttt{#1}}\fi
\expandafter\ifx\csname urlprefix\endcsname\relax\def\urlprefix{URL }\fi
\expandafter\ifx\csname href\endcsname\relax
  \def\href#1#2{#2} \def\path#1{#1}\fi

\bibitem{lu2009toward}
T.~Lu, C.~K. Law, Toward accommodating realistic fuel chemistry in large-scale computations, Prog. Energy Combust. Sci. 35 (2009) 192--215.

\bibitem{trendsturbulent}
T.~C. M. A.~N. Trends, Turbulent combustion modeling advances new trends and perspectives fluid mechanics and its applications.

\bibitem{pope2013small}
S.~B. Pope, Small scales, many species and the manifold challenges of turbulent combustion, Proc. Combust. Inst. 34 (2013) 1--31.

\bibitem{curran2019developing}
H.~J. Curran, Developing detailed chemical kinetic mechanisms for fuel combustion, Proc. Combust. Inst. 37 (2019) 57--81.

\bibitem{babkovskaia2011high}
N.~Babkovskaia, N.~E.~L. Haugen, A.~Brandenburg, A high-order public domain code for direct numerical simulations of turbulent combustion, J. Comput. Phys. 230 (2011) 1--12.

\bibitem{lu2005directed}
T.~Lu, C.~K. Law, A directed relation graph method for mechanism reduction, Proc. Combust. Inst. 30 (2005) 1333--1341.

\bibitem{lu2006linear}
T.~Lu, C.~K. Law, Linear time reduction of large kinetic mechanisms with directed relation graph: n-heptane and iso-octane, Combust. Flame 144 (2006) 24--36.

\bibitem{luo2010reduced}
Z.~Luo, T.~Lu, M.~J. Maciaszek, S.~Som, D.~E. Longman, A reduced mechanism for high-temperature oxidation of biodiesel surrogates, Energy Fuels 24 (2010) 6283--6293.

\bibitem{lu2006systematic}
T.~Lu, C.~K. Law, Systematic approach to obtain analytic solutions of quasi steady state species in reduced mechanisms, J. Phys. Chem. A 110 (2006) 13202--13208.

\bibitem{tosatto2011transport}
L.~Tosatto, B.~Bennett, M.~Smooke, A transport-flux-based directed relation graph method for the spatially inhomogeneous instantaneous reduction of chemical kinetic mechanisms, Combust. Flame 158 (2011) 820--835.

\bibitem{pepiot2008efficient}
P.~Pepiot-Desjardins, H.~Pitsch, An efficient error-propagation-based reduction method for large chemical kinetic mechanisms, Combust. Flame 154 (2008) 67--81.

\bibitem{tomlin2013role}
A.~S. Tomlin, The role of sensitivity and uncertainty analysis in combustion modelling, Proc. Combust. Inst. 34 (2013) 159--176.

\bibitem{vom2019sensitivity}
F.~vom Lehn, L.~Cai, H.~Pitsch, Sensitivity analysis, uncertainty quantification, and optimization for thermochemical properties in chemical kinetic combustion models, Proc. Combust. Inst. 37 (2019) 771--779.

\bibitem{nouri2022skeletal}
A.~Nouri, H.~Babaee, P.~Givi, H.~Chelliah, D.~Livescu, Skeletal model reduction with forced optimally time dependent modes, Combust. Flame 235 (2022) 111684.

\bibitem{vajda1985principal}
S.~Vajda, P.~Valko, T.~Turanyi, Principal component analysis of kinetic models, Int. J. Chem. Kinet. 17 (1985) 55--81.

\bibitem{turanyi1990reduction}
T.~Turanyi, Reduction of large reaction mechanisms, New J. Chem. 14 (1990) 795--803.

\bibitem{xu1999simplification}
M.~Xu, Y.~Fan, J.~Yuan, Simplification of the mechanism of nox formation in a ch4/air combustion system, Int. J. Energy Res. 23 (1999) 1267--1276.

\bibitem{turanyi1989reaction}
T.~Turanyi, T.~Berces, S.~Vajda, Reaction rate analysis of complex kinetic systems, Int. J. Chem. Kinet. 21 (1989) 83--99.

\bibitem{borger1992extended}
I.~B{"o}rger, A.~Merkel, J.~Lachmann, H.~Spangenberg, T.~Tur{'a}nyi, An extended kinetic model and its reduction by sensitivity analysis for the methanol/oxygen gas-phase thermolysis, Acta Chim. Hung. 129 (1992) 855--855.

\bibitem{zsely2003influence}
I.~G. Zs{'e}ly, T.~Tur{'a}nyi, The influence of thermal coupling and diffusion on the importance of reactions: The case study of hydrogen--air combustion, Phys. Chem. Chem. Phys. 5 (2003) 3622--3631.

\bibitem{bahlouli2014reduced}
K.~Bahlouli, U.~Atikol, R.~K. Saray, V.~Mohammadi, A reduced mechanism for predicting the ignition timing of a fuel blend of natural-gas and n-heptane in hcci engine, Energy Convers. Manag. 79 (2014) 85--96.

\bibitem{petzold1999model}
L.~Petzold, W.~Zhu, Model reduction for chemical kinetics: An optimization approach, AIChE J. 45 (1999) 869--886.

\bibitem{edwards2000reaction}
K.~Edwards, T.~Edgar, V.~Manousiouthakis, Reaction mechanism simplification using mixed-integer nonlinear programming, Comput. Chem. Eng. 24 (2000) 67--79.

\bibitem{androulakis2000kinetic}
I.~P. Androulakis, Kinetic mechanism reduction based on an integer programming approach, AIChE J. 46 (2000) 361--371.

\bibitem{banerjee2003development}
I.~Banerjee, M.~G. Ierapetritou, Development of an adaptive chemistry model considering micromixing effects, Chem. Eng. Sci. 58 (2003) 4537--4555.

\bibitem{mitsos2008optimal}
A.~Mitsos, G.~M. Oxberry, P.~I. Barton, W.~H. Green, Optimal automatic reaction and species elimination in kinetic mechanisms, Combust. Flame 155 (2008) 118--132.

\bibitem{elliott2005reaction}
L.~Elliott, D.~B. Ingham, A.~G. Kyne, N.~S. Mera, M.~Pourkashanian, C.~W. Wilson, Reaction mechanism reduction and optimization using genetic algorithms, Ind. Eng. Chem. Res. 44 (2005) 658--667.

\bibitem{elliott2006reaction}
L.~Elliott, D.~B. Ingham, A.~G. Kyne, N.~S. Mera, M.~Pourkashanian, S.~Whittaker, Reaction mechanism reduction and optimisation for modelling aviation fuel oxidation using standard and hybrid genetic algorithms, Comput. Chem. Eng. 30 (2006) 889--900.

\bibitem{baranwal2024spin}
M.~Baranwal, J.~C. Saldinger, D.~Kim, P.~Elvati, A.~O. Hero, A.~Violi, Spin: A data-driven model to reduce large chemical reaction networks, Fuel 367 (2024) 131299.

\bibitem{fang2025data}
S.~Fang, S.~Zhang, Z.~Li, W.~Han, Q.~Fu, C.-W. Zhou, L.~Yang, A data-driven sparse learning approach to reduce chemical reaction mechanisms, Combust. Flame 279 (2025) 114337.

\bibitem{jax2018github}
J.~Bradbury, R.~Frostig, P.~Hawkins, M.~J. Johnson, C.~Leary, D.~Maclaurin, G.~Necula, A.~Paszke, J.~Vander{P}las, S.~Wanderman-{M}ilne, Q.~Zhang, \href{http://github.com/google/jax}{{JAX}: composable transformations of {P}ython+{N}um{P}y programs} (2018).
\newline\urlprefix\url{http://github.com/google/jax}

\bibitem{gri_mech3}
Gri-mech 3.0, \url{http://www.me.berkeley.edu/gri_mech/}, gRI-Mech 3.0 Chemical Kinetic Mechanism (1999).

\bibitem{usc_mechII}
H.~Wang, X.~You, A.~V. Joshi, S.~G. Davis, A.~Laskin, F.~Egolfopoulos, C.~K. Law, Usc mech version ii: High-temperature combustion reaction model of h2/co/c1--c4 compounds, \url{http://ignis.usc.edu/USC_Mech_II.htm} (May 2007).

\bibitem{mcbride2002nasa}
B.~J. McBride, NASA Glenn coefficients for calculating thermodynamic properties of individual species, National Aeronautics and Space Administration, John H. Glenn Research Center, 2002.

\bibitem{kingma2014adam}
D.~P. Kingma, Adam: A method for stochastic optimization, arXiv preprint arXiv:1412.6980 (2014).

\bibitem{yang2019quantization}
J.~Yang, X.~Shen, J.~Xing, X.~Tian, H.~Li, B.~Deng, J.~Huang, X.-s. Hua, Quantization networks, in: Proc. IEEE/CVF Conf. Comput. Vis. Pattern Recognit., 2019, pp. 7308--7316.

\end{thebibliography}


\newpage

\small


\end{spacing}
\end{document}